\documentclass[aps,prd,showpacs,nofootinbib,preprintnumbers,twocolumn]{revtex4-1}
\usepackage{amsmath,amssymb}
\usepackage[dvips]{graphicx}
\usepackage[bookmarks=true,bookmarksnumbered=true,
colorlinks=false,pdftitle={Paper},pdfauthor={Hiroaki Abuki, Gordon Baym, 
Tetsuo Hatsuda, Naoki Yamamoto},
pdfsubject={},pdfkeywords={}]{hyperref}
\usepackage{enumerate}
\newcommand \beq{\begin{eqnarray}}
\newcommand \eeq{\end{eqnarray}}

\newcommand{\SU}{\text{SU}}
\newcommand{\U}{\text{U}}

\DeclareMathOperator{\tr}{Tr}

\def\simge{\mathrel{
       \rlap{\raise 0.511ex \hbox{$>$}}{\lower 0.511ex \hbox{$\sim$}}}}
\def\simle{\mathrel{
       \rlap{\raise 0.511ex \hbox{$<$}}{\lower 0.511ex \hbox{$\sim$}}}}

\begin{document}
\preprint{TKYNT-10-02}
\title{The NJL model of dense three-flavor matter with axial anomaly:\\
 the low temperature critical point and BEC-BCS diquark crossover }

\author{Hiroaki Abuki$^1$}
%\email{h.abuki@rs.kagu.tus.ac.jp}
\author{Gordon Baym$^2$}
%\email{gbaym@uiuc.edu}
\author{Tetsuo Hatsuda$^3$}
%\email{hatsuda@phys.s.u-tokyo.ac.jp}
\author{Naoki Yamamoto$^3$}
%\email{yamamoto@nt.phys.s.u-tokyo.ac.jp}
\affiliation{$^1$Department of Physics, Tokyo University of Science,
Tokyo 162-8601, Japan}
\affiliation{$^2$Department of Physics, University of Illinois, 1110 W. Green St.,
Urbana, Illinois 61801, USA, } 
\affiliation{$^3$Department of Physics, The University of Tokyo, Tokyo
  113-0033, Japan}
\date{\today}

\begin{abstract}
We study the QCD phase structure in the three-flavor Nambu--Jona-Lasinio 
model, incorporating the interplay between the chiral and diquark condensates 
induced by the axial anomaly. We demonstrate that for an appropriate range
of parameters of the model, the interplay leads to the low temperature critical 
point in the phase structure predicted by a previous Ginzburg-Landau analysis. 
We also show that a Bose-Einstein condensate (BEC) of diquark molecules emerges 
in the intermediate density region, and as a result, a BEC-BCS crossover is 
realized with increasing quark chemical potential.
\end{abstract}

\maketitle

\section{Introduction}

The phases of strongly interacting matter described by quantum chromodynamics 
(QCD) at finite temperature $T$ and quark chemical potential $\mu$ is being 
actively studied theoretically, as well as experimentally in ultrarelativistic 
heavy ion collisions at RHIC (Relativistic Heavy Ion Collider) and in the near 
future at the LHC (Large Hadron Collider).   
At low $T$ and $\mu$, the hadronic phase is realized with chiral symmetry 
dynamically broken by condensation of quark-antiquark pairs, the chiral 
condensate $\langle \bar q q \rangle$. 
 On the other hand,  at low $T$ and high $\mu$ a color superconducting (CSC) 
phase \cite{Alford2008},  characterized by formation of quark-quark pairs 
-- a diquark condensate $\langle qq \rangle$ -- is expected to appear owing to 
the attractive one-gluon exchange interaction or the instanton-induced 
interaction in the quark-quark channel. 
At high $T$ for any $\mu$, the quark-gluon plasma (QGP) phase \cite{Yagi2005} 
is realized with both the chiral and diquark condensates melted away. 
The phase transition from the hadronic phase to the QGP phase is indeed 
confirmed by recent lattice QCD Monte Carlo simulations indicating a smooth 
crossover for physical quark masses \cite{Aoki2009, Bazavov2009}.

Nevertheless, the first-principles lattice technique based on importance sampling
is not applicable to QCD at finite $\mu$ due to the complex fermion determinant.
This is why our understanding of the transition from the hadronic phase 
to the CSC phase relevant to the compact star physics is still immature 
and we have to basically rely on specific models of QCD, such as the 
Nambu--Jona-Lasinio (NJL) model \cite{Hatsuda1994, Buballa2005}, 
the Polyakov--Nambu--Jona-Lasinio (PNJL) model 
\cite{Fukushima2004a,*Fukushima2008b, Ratti2006, Roessner2007, Hell2009}, and 
the random matrix theory (RMT) \cite{Halasz1998a, Vanderheyden2000a, Sano2009}.
These model studies together with the lattice QCD results have revealed 
the possible existence of the critical point \cite{Asakawa1989, Barducci1989}
at high $T$ between the hadronic phase and the QGP phase 
(see, however, \cite{Forcrand2008}).

\begin{figure}[h]
\begin{center}
\includegraphics[width=6cm]{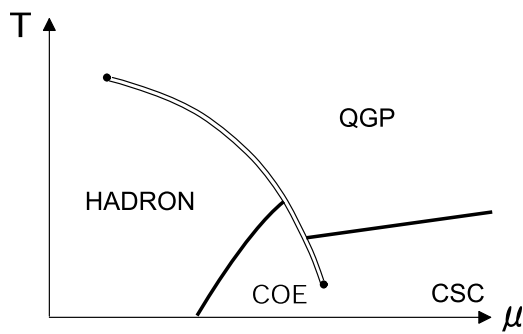}
\end{center}
\vspace{-0.5cm}
\caption{Schematic phase structure with two light (up and down) quarks and
a medium heavy (strange) quark.  In the hadronic phase, $q\bar{q}$ pairs 
condense, while in the color superconducting (CSC) phase, the dominant 
condensation is $qq$ pairing. In the quark-gluon plasma (QGP), all symmetries 
are restored without any pairing, while in the coexistence (COE) region 
$q\bar{q}$ and $qq$ pairings coexist. The double line denotes a first-order 
phase transition. Adapted from Ref.~\cite{Yamamoto2007a}.
}
\label{fig:32F}
\end{figure}

Recently, we have pointed out the possibility of a new low temperature 
critical point between the hadron phase and the CSC phase in three-flavor 
QCD, on the basis of model-independent Ginzburg-Landau (GL) theory 
\cite{Hatsuda:2006ps, Yamamoto2007a}: the attraction between the chiral 
and diquark condensates induced by the axial anomaly leads to this critical 
point and an associated smooth crossover.  
Figure~{\ref{fig:32F}} illustrates the dense three-flavor QCD phase diagram 
with the new critical point at low $T$ \cite{Yamamoto2007a}.
This may provide a mechanism of continuity between hadronic matter and 
quark matter (hadron-quark continuity) conjectured by Sch\"{a}fer and Wilczek 
\cite{Schafer:1998ef}.
Moreover, the idea of hadron-quark continuity is corroborated by recent studies 
on the spectral continuity of Nambu-Goldstone modes \cite{Yamamoto2007a} 
and vector mesons \cite{Hatsuda2008}, and the formal similarity of the partition 
functions in the universal regime between the low and high $\mu$ regimes in 
three-flavor QCD at finite size, large compared with the inverse gap, but small 
compared with the pion Compton wavelength {\cite{Yamamoto2009}.
In two-flavor QCD, similar new critical points have also been found in the
NJL model \cite{Kitazawa2002a, *Kitazawa2003, Zhang2009, Zhang2009a},
although their origin is related to a repulsive vector-channel four-fermion 
interaction \cite{Kitazawa2002a,*Kitazawa2003}, or electric charge neutrality 
and $\beta$-equilibrium conditions \cite{Zhang2009, Zhang2009a} rather than 
the axial anomaly. These studies may imply smooth crossovers not only as a function 
of $T$ at low $\mu$ but also as a function of $\mu$ at low $T$
in the realistic QCD phase diagram.

The Ginzburg-Landau analysis for three-flavor QCD in 
Refs.~\cite{Hatsuda:2006ps, Yamamoto2007a} depends on the assumption that 
the magnitudes of the chiral and diquark condensates are sufficiently small 
near the phase boundaries, which may not be justified over the entire region 
in the QCD phase diagram and for strongly first-order phase transitions. 
The question is unanswered as to whether such a new critical point
induced by the axial anomaly does really emerge in the ($\mu,T$)-plane 
within the framework of phenomenological models with reasonable parameters.

The purposes of this paper are two-fold.   First, using the three-flavor 
NJL model incorporating the interplay between the chiral and diquark 
condensates induced by the axial anomaly, we study the location of the 
new critical point predicted in \cite{Hatsuda:2006ps, Yamamoto2007a}. 
We demonstrate that this critical point indeed appears in the phase 
diagram for an appropriate range of parameters.
Second, we show that the axial anomaly also triggers, in this model, a crossover
between a Bose-Einstein condensed state (BEC) of diquark pairing
and Bardeen-Cooper-Schrieffer (BCS) diquark pairing.
This BEC-BCS crossover is identical in structure to that 
in nonrelativistic condensed matter systems \cite{Eagles1969a, Leggett1980, Nozieres1985}; 
and is discussed for relativistic systems \cite{Abuki2002, Nishida2005, Nawa2006, 
Abuki2007, Deng2007, Sun2007, He2007, Kitazawa2008, Brauner2008, Baym2008}:
the change in size of Cooper pairs at lower $\mu$ within the QCD Schwinger-Dyson approach 
\cite{Abuki2002}, the relativistic BEC-BCS crossover of diquark pairing in the NJL-type model 
\cite{Nishida2005, Abuki2007, Sun2007, He2007, Kitazawa2008, Brauner2008} and
the diquark-quark model \cite{Nawa2006, Deng2007}, and a possible evolution from
baryons in nuclear matter to diquarks in quark matter with increasing $\mu$ \cite{Baym2008} 
are elucidated. Remarkably, as we will show in this paper, in relativistic quark matter 
at high $\mu$ the axial anomaly enhances the attractive interaction between quarks, 
leading to the emergence of a BEC state of diquark pairing.

To illustrate the essential physics induced by the axial anomaly and to avoid 
complications of charge neutrality and $\beta$-equilibrium conditions, we assume 
$\SU(3)$ flavor symmetry $m_u=m_d=m_s \equiv m_q$ throughout this paper. 
The generalization to include these effects will be reported elsewhere.

This paper is organized as follows. In Sec.~\ref{sec:njl}, we
formulate the three-flavor NJL model incorporating the interplay
between the chiral and diquark condensates induced by the axial anomaly.
In Sec.~\ref{sec:phase}, we discuss the phase structures
with and without the interplay.
In Sec.~\ref{sec:bec-bcs}, we show that the interplay leads not only to the new 
critical point but also to the BEC-BCS crossover of the diquark pairing 
at high density. Sec.~\ref{sec:summary} is devoted to a summary and 
concluding remarks.

\section{NJL model with axial anomaly}
\label{sec:njl}
The Lagrangian of the Nambu--Jona-Lasinio (NJL) model with three-flavors
consists of three terms:
\beq
\label{eq:njl}
{\cal L}=
\bar q (i\gamma_{\mu}\partial^{\mu}-m_q  + \mu \gamma_0) q 
+ {\cal L}^{(4)} + {\cal L}^{(6)},
\eeq
where $q=({\rm u,d,s})^T$ is transpose of the quark field, 
$m_q$ is a flavor symmetric quark mass ($m_u=m_d=m_s$).
${\cal L}^{(4)}$ and ${\cal L}^{(6)}$ are the 
four-fermion interaction and six-fermion interaction, respectively.
The standard choice of ${\cal L}^{(4)}$ is \cite{Hatsuda1994, Buballa2005},
\beq
{\cal L}^{(4)}&=&{\cal L}^{(4)}_{\chi} + {\cal L}^{(4)}_{d}, \\ 
{\cal L}^{(4)}_{\chi}&=& G \sum_{a=0}^8 
\left[(\bar q \tau_a q)^2 + (\bar q i \gamma_5 \tau_a q)^2 \right] \nonumber \\
 &=& 8 G {\rm tr} ({\phi^{\dag} \phi}), \\
{\cal L}^{(4)}_{d}&=& H \! \! \sum_{A, A'=2,5,7} 
\left[(\bar q i \gamma_5 \tau_A \lambda_{A'} C \bar q^T)
(q^T C i \gamma_5 \tau_A \lambda_{A'}q) \right. \nonumber \\
& & \left. \ \ \ \ \ \ \ \ \ \ \ \ \ \ \ \ 
+(\bar q \tau_A \lambda_{A'} C \bar q^T)
(q^T C \tau_A \lambda_{A'}q)\right]   \nonumber \\
 &=& 2H {\rm tr} [d_L^{\dag}d_L + d_R^{\dag}d_R],
 \label{eq:qq}
\eeq
where $\phi_{ij} \equiv (\bar{q}_R)^j_a (q_L)^i_a$,
$(d_L)_{ai} \equiv \epsilon_{abc} \epsilon_{ijk} ({q}_L)^j_b C (q_L)^k_c$,
and $(d_R)_{ai} \equiv \epsilon_{abc} \epsilon_{ijk} ({q}_R)^j_b C (q_R)^k_c$,
with $a, b, c$ and $i, j, k$  the color and flavor indices,
and $C$ the charge conjugation operator. ${\rm tr}$ is taken over the flavor indices.
 The flavor $\U(3)$ generators $\tau_a$ ($a=0, \cdots, 8$) are normalized
so that ${\rm tr}[\tau_a \tau_b]=2\delta_{ab}$, and $\tau_A$ and $\lambda_{A'}$ 
with $A,A'=2,5,7$ are antisymmetric generators of flavor and $\SU(3)$ color, respectively.
The coupling constants $G$ and $H$ with dimension $(\rm mass)^{-2}$ are assumed
to be positive.  Starting from the one-gluon exchange interaction and
apply a simple Fierz transformation, we obtain the ratio
$H/G=3/4$.   However, we treat $G$ and $H$ as independent parameters of the 
effective Lagrangian and as detailed in Sec.~\ref{sec:phase} take the values common in the 
literature.

The four-fermion interactions introduced above are invariant under
$\SU(3)_L \times \SU(3)_R \times \U(1)_A \times \U(1)_B$ symmetry.  
The interaction ${\cal L}^{(4)}_{\chi}$ produces attraction of $q\bar{q}$ pairs in the  
color-singlet and spin-parity $0^{\pm}$ channel, 
inducing dynamical breaking of chiral symmetry with formation 
of a chiral condensate \cite{Nambu1961,*Nambu1961a}.
Similarly ${\cal L}^{(4)}_{d}$ leads to attraction of $qq$ pairs in the  
color-anti-triplet and spin-parity $0^{\pm}$ channel,  inducing color-flavor locked (CFL)
superconductivity with formation of a diquark condensate 
\cite{Alford:1998mk}.

The six-fermion interaction in our model consists of two parts,
\beq
{\cal L}^{(6)}={\cal L}^{(6)}_{\chi}+{\cal L}^{(6)}_{\chi d}.
\eeq
${\cal L}^{(6)}_{\chi}$ is the standard
Kobayashi-Maskawa-'t Hooft (KMT) interaction
\cite{Kobayashi1970,Hooft1976,*Hooft1976a}, 
\beq
\label{eq:kmt}
{\cal L}^{(6)}_{\chi}=- 8 K 
\left(\det \phi +{\rm h.c.}\right).
\eeq
This interaction, invariant under $\SU(3)_L \times \SU(3)_R 
\times \U(1)_B$ symmetry but not under $\U(1)_A$ symmetry, 
accounts for the axial anomaly in QCD due to instantons.
For positive coupling constant $K$ with dimension $({\rm mass})^{-5}$, as we assume, 
the $\eta'$ meson has a larger mass 
than the other pseudo Nambu-Goldstone bosons ($\pi, \eta, K$).
The term (\ref{eq:kmt}) serves the role in the QCD phase structure of making
 the chiral phase transition first-order as a function of $T$ 
at $\mu=0$ for massless three-flavor limit \cite{Pisarski1984}.

As pointed out in \cite{Hatsuda:2006ps,Yamamoto2007a}, 
the instanton couples {\rm the diquark condensate and the chiral condensate}, 
which modifies the QCD phase structure in the intermediate density region.
The effective interaction between the chiral and diquark pairing fields
is described by a six-fermion term,
\beq
\label{eq:K'}
{\cal L}^{(6)}_{\chi d} 
= K' \left(\tr[(d_R^{\dag} d_L) \phi] +{\rm h.c.} \right),
\eeq
which has $\SU(3)_L \times \SU(3)_R \times \U(1)_B$ symmetry but breaks
$\U(1)_A$ symmetry explicitly.  It is this term that is responsible for the low 
temperature critical point.
We assume $K'>0$, so that $qq$ pairs in the positive parity channel, 
$\langle d_L \rangle =-\langle d_R \rangle$, are energetically favored,
as suggested from the weak-coupling instanton calculations 
\cite{Schafer2002a, Yamamoto:2008zw}.  Since the term (\ref{eq:K'}) acts as 
an external field for $\chi$, it washes out the first-order chiral phase transition
at intermediate density for sufficiently large $K'|\langle d_R \rangle|^2$
\cite{Hatsuda:2006ps,Yamamoto2007a}.
If we start from the instanton vertex and apply a simple Fierz transformation, 
we obtain the ratio $K'/K=1$.  However, since there is no a priori reason 
that $K$ and $K'$ have this ratio in the effective Lagrangian level, 
we keep them as independent parameters.

The favorable condensates by the interaction
 ${\cal L}^{(4)}+{\cal L}^{(6)}$  are the 
flavor-symmetric chiral and 
diquark condensates in the spin-parity $0^{+}$ channel, defined by
\beq
\chi \delta_{ij} &=& \langle \bar q_a^i q_a^{j} \rangle, \\
s \delta_{AA'} &=& \langle q^T C \gamma_5 \tau_A \lambda_{A'} q \rangle.
\label{eq:s}
\eeq
Here the condensate order parameters $\chi$ and $s$, which  are 
proportional to  the order parameters $\sigma$ and $d$  
defined in the previous Ginzburg-Landau analysis \cite{Hatsuda:2006ps, Yamamoto2007a},
 are related to the parameters
 $\phi$ and $d_{L,R}$ defined below Eq.~(\ref{eq:qq}) by
\beq
\chi \delta_{ij} &=& 2 \langle \phi_{ij} \rangle, \\
s \delta_{ai} &=& 2\langle (d_L)_{ai}
 \rangle=-2\langle (d_R)_{ai} \rangle. 
\eeq

 We work at the mean-field level, linearizing the 
 the products of operators $X$ and $Y$ as
  $X^2 \rightarrow 2 \langle X \rangle X - \langle X \rangle^2$, 
  $XY \rightarrow \langle X \rangle Y + \langle Y \rangle X 
  - \langle X \rangle \langle Y \rangle$, and 
   $X^2Y \rightarrow \langle X \rangle^2 Y + 2 \langle X \rangle 
   \langle Y \rangle X  - 2 \langle X \rangle^2 \langle Y \rangle$.
Subtraction of the constant terms 
avoids double counting the interactions. 
In mean-field deviations from factorization are partially compensated 
for by redefinition of the coupling constants $G$, $H$, $K$, and $K'$.  
Then ${\cal L}^{(4)}$ and ${\cal L}^{(6)}$ reduce to
\beq
{\cal L}^{(4)}_{\chi} &\rightarrow & 4G \chi \bar q q - 6G \chi^2, 
\nonumber \\
{\cal L}^{(4)}_{d} & \rightarrow  & 
H\left[s^*(q^T C\gamma_5 \tau_A \lambda_{A} q) + {\rm h.c.} \right] - 3 H |s|^2,
\nonumber \\
{\cal L}^{(6)}_{\chi} & \rightarrow & -2K \chi^2 \bar q q + 4K \chi^3, 
\nonumber \\
{\cal L}^{(6)}_{\chi d}&\rightarrow&
-\frac{K'}{4}|s|^2 \bar q q
-\frac{K'}{4} \chi \left[s^* (q^T C\gamma_5 \tau_A \lambda_{A} q) + {\rm h.c.}\right]
\nonumber \\
& &+\frac{3K'}{2}|s|^2 \chi.
\eeq
Here and below we implicitly sum over $A=2,5,7$ unless otherwise stated.

To derive the thermodynamic potential, it is most convenient to work in the 
Nambu-Gor'kov formalism; we introduce the bispinor field 
\beq
\Psi=\frac{1}{\sqrt{2}}(q, q^C)^T,
\eeq
with $q^C=C \bar q ^T$ (and $\bar q^C = q^T C$) the charge-conjugate quark field.
Then the linearized form of the NJL Lagrangian becomes
\beq
\label{eq:bispinor}
{\cal L} = \bar \Psi S^{-1} \Psi - U.
\eeq
Here $S^{-1}(p)$ is the inverse propagator in the momentum space:
\beq
\label{eq:propagator}
\! \! \! \! \! S^{ - 1}(p)  = \left( {\begin{array}{*{20}c}
\! \gamma_{\mu}p^{\mu} + \mu \gamma_0 - M 
& \! \Delta \gamma_5 \tau_A \lambda_{A} \\
\! { - \Delta ^* }\gamma_5 \tau_A \lambda_{A} 
& \! \gamma_{\mu}p^{\mu} - \mu \gamma_0 - M \\
\end{array}} \right),
\eeq
where the dynamical Dirac mass in the $q\bar{q}$-channel reads
\beq
M(\chi,s,m_q) = m_q-4 \left(G-\frac{1}{8}K \chi \right) \chi + \frac{1}{4}{K'}|s|^2,
\label{eq:M}
\eeq
and the dynamical Majorana mass in the $qq$-channel reads
\beq
\Delta(\chi,s) = -2 \left(H - \frac{1}{4} K' \chi \right) s.
\label{eq:Delta}
\eeq
They are both dependent on the order parameters, $\chi$ and $s$.
 The constant term needed to subtract double counting of the interactions
 in ${\cal L}$ is
\beq
U(\chi,s) = 6G \chi^2 + 3H |s|^2 - 4K \chi^3 
- \frac{3}{2} K' |s|^2 \chi .
\label{eq:V}
\eeq
The terms in Eqs.~(\ref{eq:M}), (\ref{eq:Delta}),
and (\ref{eq:V}) are shown diagrammatically in 
Figs.~\ref{fig:Dirac}(a)-(d), \ref{fig:Majorana}(a)-(b), 
and \ref{fig:potential}(a)-(d), respectively.  The chiral-diquark 
coupling (the $K'$-term)
enhances the attractions in both the $\bar q q$ and $qq$ 
channels.

\begin{figure}[t]
\begin{center}
\includegraphics[width=6cm]{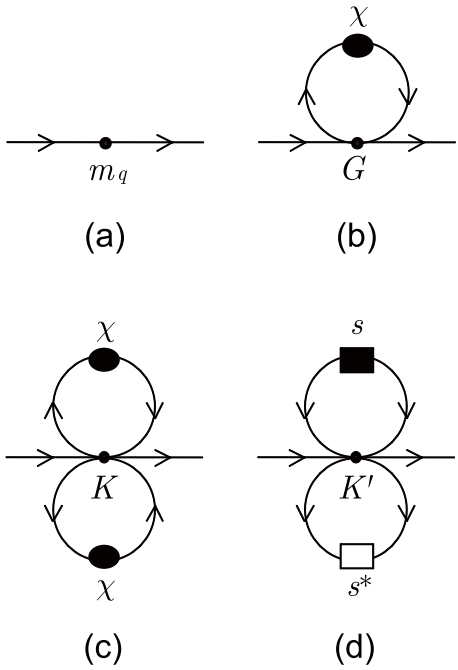}
\end{center}
\vspace{-0.5cm}
\caption{Four contributions to the Dirac mass $M$ (constituent quark mass).
The chiral condensate $\chi$ is denoted by a black circle, the 
diquark condensate $s$ by a black square, and
$s^*$ by a white square.}
\label{fig:Dirac}
\end{figure}

\begin{figure}[h]
\begin{center}
\includegraphics[width=6cm]{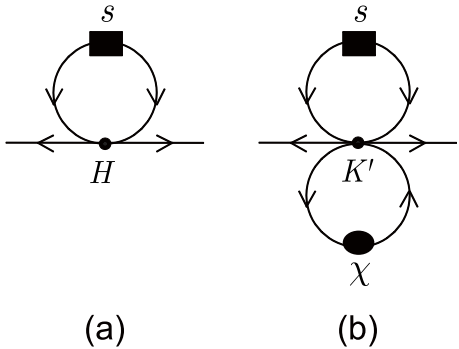}
\end{center}
\vspace{-0.5cm}
\caption{Two contributions to the Majorana mass $\Delta$ (superconducting gap).
}
\label{fig:Majorana}
\end{figure}

\begin{figure}[h]
\begin{center}
\includegraphics[width=7cm]{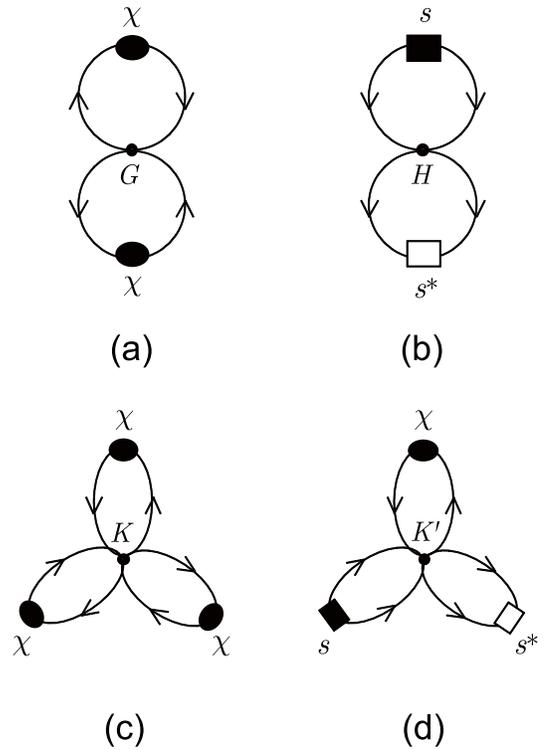}
\end{center}
\vspace{-0.5cm}
\caption{Four contributions to the constant term $U$.
}
\label{fig:potential}
\end{figure}

The thermodynamic potential at temperature $T$ and quark chemical potential $\mu$ 
is given by
\beq
\! \! \! \! \! \! \! \! \!
\Omega \! = \! -T\sum_n \! \int \! \frac{d^3p}{(2\pi)^3}\frac{1}{2}{\bf {\tr}} \ln \!
\left[\frac{1}{T} S^{-1}(i\omega_n, \vec p) \right] \!+\! U(\chi,s),
\eeq
where $\tr$ is taken over the bispinor space with 
  the factor 1/2 in front to correct for double counting of
degrees of freedom.
Evaluating the trace and summing over 
the fermionic Matsubara frequencies $p^0=i\omega_n=(2n+1)\pi  iT$, we
arrive at the thermodynamic potential  \cite{Buballa2005}:
\beq
\label{eq:thermo-pot}
\Omega &=&
-\int \frac{d^3p}{(2\pi)^3}\sum_{\pm}
\biggl\{ 
\left[16T \ln(1+e^{-\omega_8^{\pm}/T}) +8 \omega_8^{\pm} \right]
\nonumber \\
& & \ \ \ 
+\left[2T \ln(1+e^{-\omega_1^{\pm}/T}) + \omega_1^{\pm} \right] \biggr\}
+U(\chi,s),
\eeq
where
\beq
\label{eq:dispersion1}
\omega_8^{\pm}&=&\sqrt{(E_p \pm \mu)^2+ \Delta_1^2}, 
\\
\label{eq:dispersion2}
\omega_1^{\pm}&=&\sqrt{(E_p \pm \mu)^2+ \Delta_8^2},
\eeq
are the dispersion relations for the quasiquarks in the octet and singlet
representations, with $E_p=\sqrt{p^2+ M^2}$, $\Delta_1=2\Delta$,
and $\Delta_8 = \Delta$.
Equations (\ref{eq:M}) and (\ref{eq:thermo-pot}) imply that
$\chi<0$ is energetically favored for non-zero $m_q$.
On the other hand, $s$ is generally complex and 
the thermodynamic potential is a function of $|s|^2$.

\section{Phase structure}
\label{sec:phase}
We now explore the 
effect of the attractive $K'$-term induced by the axial anomaly on the
phase structure in the ($\mu,T$)-plane of the three-flavor NJL model. 
The phase structures can be determined numerically
by looking for the values of $\chi$ and $s$ that minimize  the thermodynamic potential 
in Eq.~(\ref{eq:thermo-pot}) globally. 
We follow the parameter choice of \cite{Buballa2005} where the
coupling constants $G$ and $K$ are chosen to fit empirical 
mesonic quantities and the chiral condensate in the QCD vacuum.
Table \ref{tab} shows two sets of parameters we adopt below.
We vary the strength of the  chiral-diquark coupling (the $K'$ term) by hand.
In order to illustrate how the anomaly changes the 
conventional phase structure and to avoid the complications of charge neutrality 
and $\beta$-equilibrium, we assume $\SU(3)$ flavor symmetry, $m_u=m_d=m_s \equiv m_q$.
\begin{table}[t]
\begin{tabular}{|c|c|c|c|c|c|c|}
\hline 
& $m_q$ [MeV] & \ $G\Lambda^2$ \ & \ $H\Lambda^2$ \
& \ $K\Lambda^5$ \ & $M$ [MeV] & $\chi^{1/3}$ [MeV] \\ \hline \hline
\ I \ & 0 & 1.926 & 1.74 & 12.36 & 355.2 & $-$240.4 \\ \hline
\ II \ & 5.5 & 1.918 & 1.74 & 12.36 & 367.6 & $-$241.9  \\  \hline
\end{tabular}
\caption{Two sets of parameters in the present three-flavor NJL model: 
the current quark mass $m_q$, coupling constants $G$, $H$, 
and $K$, with a spatial momentum cutoff $\Lambda=602.3$ MeV \cite{Buballa2005}.
The dynamical quark mass $M$ and the chiral condensate $\chi$ in the vacuum 
are also given. }
\label{tab}
\end{table}

\subsection{Without the chiral-diquark interplay}

We first show the phase structures without the $K'$-term in Fig.~\ref{fig:njl0}.
Panels (a) and (b) show the results of the case of massless quarks, I, 
and finite mass quarks, II, respectively.
The phase diagram contains of a normal (NOR) phase defined by $\chi=s=0$, 
a Nambu-Goldstone (NG) phase defined by $\chi \neq 0$ and $s=0$, and
a color superconducting (CSC) phase defined by $\chi=0$ and $s \neq 0$.\footnote{
Even when chiral symmetry is broken only slightly ($\chi \sim 0$)
by the current quark mass, we  use the same 
 classification in terms of NOR, NG and COE  as  in Fig.~\ref{fig:njl0}(b).}
 The chiral phase transition between the NG and NOR (or NG and CSC) phases
is first-order, with the chiral condensate $\chi$ changing discontinuously,
while the color superconducting phase transition between the CSC and NOR phases
is second-order, with the diquark condensate $s$ changing continuously but 
not smoothly with a discontinuity in the diquark susceptibility $\partial s/\partial T$.

\begin{figure}[t]
\begin{center}
\includegraphics[width=8cm]{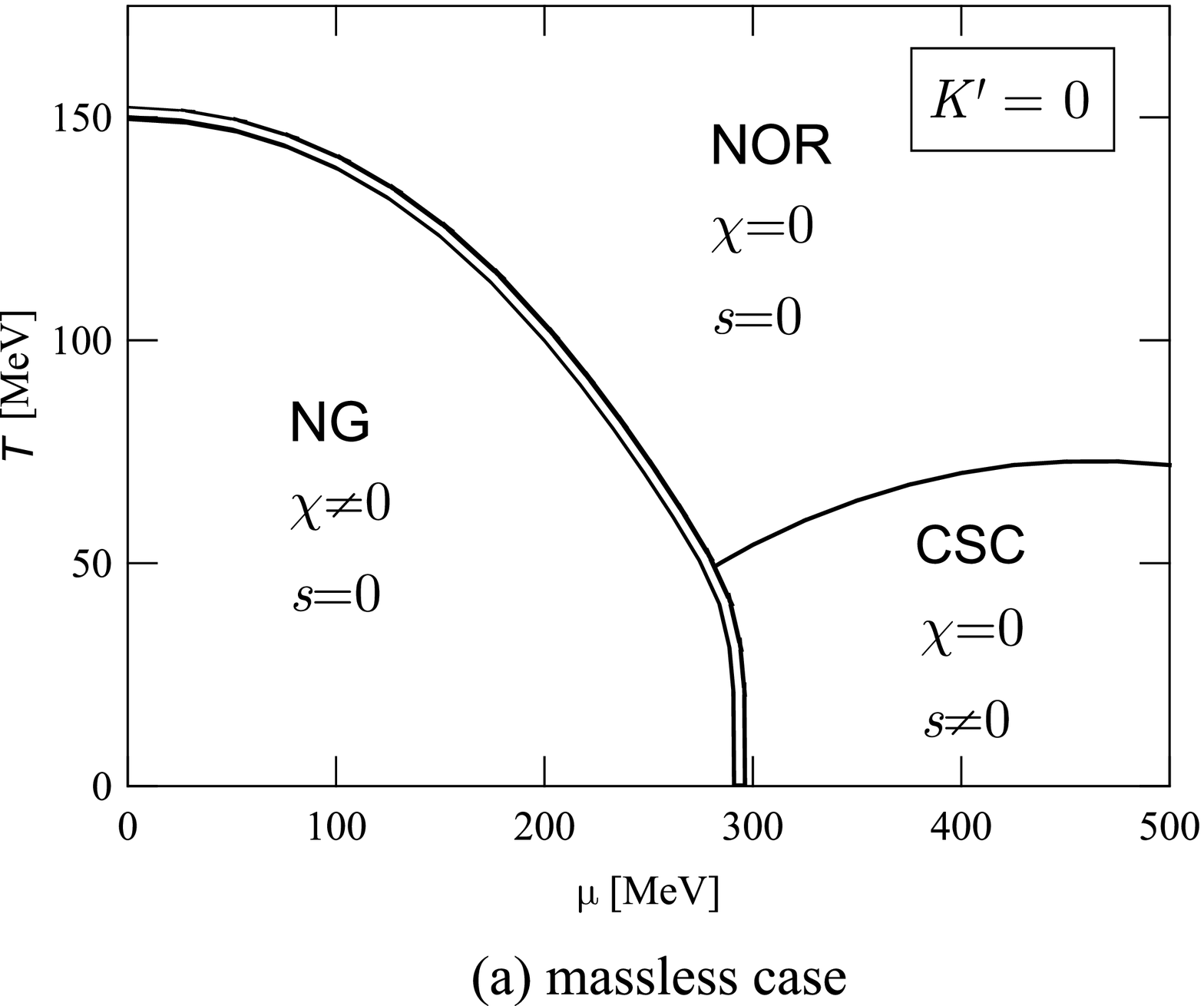}

\vspace{1cm}

\includegraphics[width=8cm]{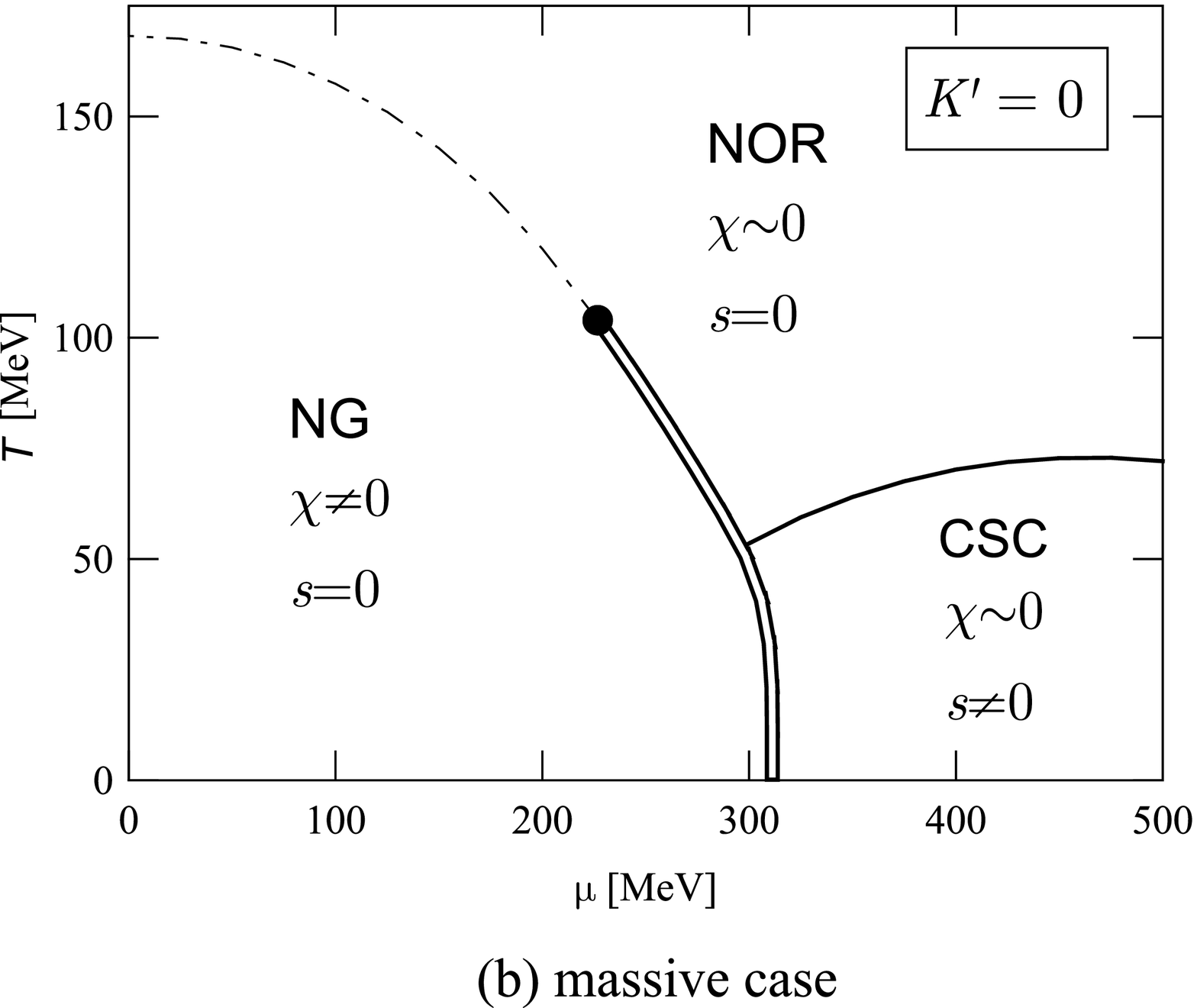}   
\caption{The phase structure in the ($\mu,T$)-plane in
the three-flavor NJL model without the axial anomaly for (a) case I, 
massless quarks, and (b) case II, finite mass quarks. 
Phase boundaries with a second-order transition are denoted 
by a single line and a first-order transition by a double line. 
The dashed-dot line at high T in panel (b) shows the effective 
chiral crossover line, at which the susceptibility $\partial \chi/\partial T$ peaks. 
See the text for further detail.}
\label{fig:njl0}
\end{center}
\end{figure}

In case II, the current quark mass changes
 first-order chiral phase transition to a crossover at high temperature, 
whereas 
 the first-order transition  at high density region still remains,
  as shown in Fig.~\ref{fig:njl0}(b).
As a result, the second-order critical point,  the Asakawa-Yazaki point 
 \cite{Asakawa1989, Barducci1989}, 
appears in the ($\mu,T$)-plane. 
The QCD critical point moves down towards 
the $\mu$-axis with increasing quark mass $m_q$.
The region $\chi \sim 0$ is characterized by explicit breaking of chiral 
symmetry by the quark mass, while in the region $\chi \neq 0$, chiral 
symmetry is dynamically broken.

\subsection{With the chiral-diquark interplay}

When the strength of the chiral-diquark coupling due to axial anomaly,
 $K'$, is relatively small ($K'< 4.1 K $ in case I
and $K'< 3.8 K$ in case II),
the topologies of the phase structures remain unchanged,  as one sees in  
 Figs.~\ref{fig:njl0}(a) and (b). 
On the other hand, once $K'$ exceeds a critical value ${K'}_c$,
 the topological structure of the phase diagram changes 
 as seen in Fig.~\ref{fig:njl1} (shown for $K'=4.2K$):
as discussed in \cite{Hatsuda:2006ps, Yamamoto2007a} 
using the Ginzburg-Landau approach, the $K'$-term, 
which acts as an external field for $\chi$,
turns the first-order chiral phase transition into a crossover,
and leads to a low $T$  critical point at intermediate density. 
As a result, the coexistence (COE) phase defined by $\chi \neq 0$ and $s \neq 0$ 
spreads over the higher density region across the second-order
phase boundary from the NG phase in both cases I and II.
The emergence of the COE phase is consistent with the model-independent result 
that the chiral condensate $\chi$ is proportional to the instanton density 
(or the strength of the axial anomaly) in the CFL phase \cite{Yamamoto:2008zw}.

In Fig.~\ref{fig:KKprime}, we depict $K'_c$ as a function of $K$ for several 
values of the current quark mass, $m_q=0$, $m_q=5.5$ MeV, and $m_q=140.7$ MeV. 
The $K'_c$-line separates the crossover and first-order regions;
the chiral-diquark coupling $K'$ favors the crossover, 
while the triple chiral coupling $K$ favors first-order.
As $m_q$ increases, the crossover region is enlarged since the current quark mass acts 
as an external field on the chiral condensate, weakening the chiral transition. 

\begin{figure}[t]
\begin{center}
\includegraphics[width=8cm]{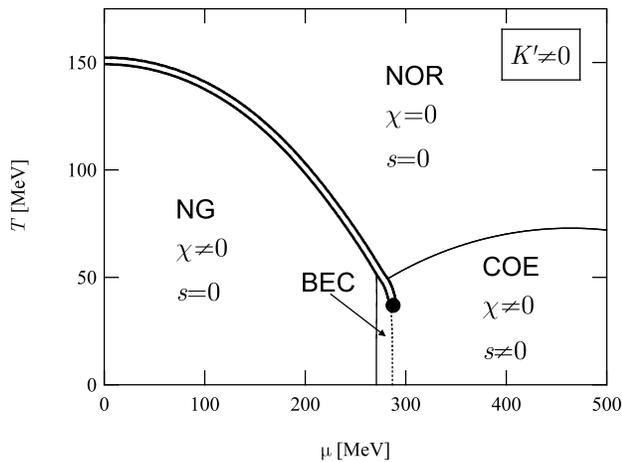}  

\vspace{1cm}

\includegraphics[width=8cm]{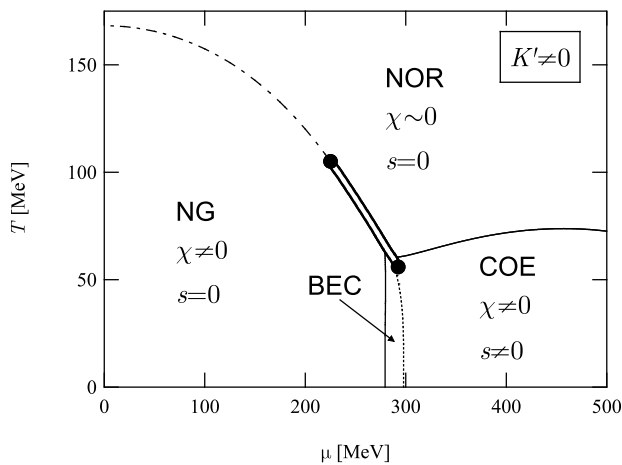} 
\caption{Phase structure in the ($\mu,T$)-plane in the three-flavor NJL model 
with the axial anomaly for (a) massless quarks, and (b) finite mass quarks. 
 The phase boundaries with a second-order transition are denoted 
by a single line and  a first-order transition by a double line. 
The BEC-BCS crossover (dotted) line in (a) and (b) is defined by $\mu=M(\mu,T)$,
 the dynamical quark mass.}
\label{fig:njl1}
\end{center}
\end{figure}

\begin{figure}[t]
\begin{center}
\includegraphics[width=7.2cm]{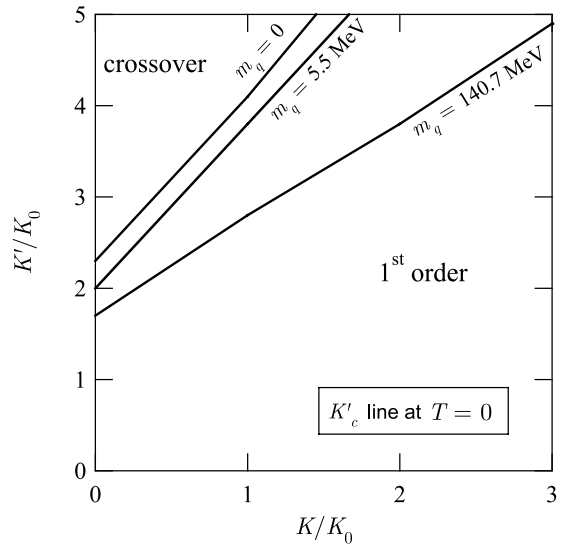}
\caption{
Critical lines in the $(K,K')$-plane at $T=0$ for several values of 
the current quark mass $m_q$ ($K_0 \Lambda^5 \equiv 12.36$). 
Chiral phase transition is realized as a smooth crossover in the region
above the corresponding line while it is of first-order below the line.
}
\label{fig:KKprime}
\end{center}
\end{figure}

\section{BEC-BCS crossover induced by the axial anomaly}
\label{sec:bec-bcs}

The axial anomaly,  for sufficiently large 
chiral-diquark coupling $K'$, not only triggers the low $T$  critical point, 
but also a BEC-BCS crossover in the COE phase, as discussed in \cite{Baym2008} 
in qualitative analogy with the cold atomic gases in condensed-matter physics.  
Physically the BEC regime is characterized by quark-pair sizes small compared 
to the interparticle spacing, while in the BCS regime the pair size is large 
compared with the interparticle spacing. 
The possibility of a BEC-BCS crossover in a color superconductor, 
and the presence of a BEC regime, was first pointed out in \cite{Abuki2002} 
by looking at the change in size of the pairs with density.  
As shown later within an NJL-type model such a BEC regime appears for sufficiently 
large pairing attraction, $H$, in the $qq$-channel \cite{Kitazawa2008}. 
The novel feature we stress here is that the axial anomaly helps to realize the BEC regime 
through its contribution to the effective $qq$ coupling in (\ref{eq:Delta}),
\beq
H' \equiv H+ \frac{1}{4}K'|\chi|.
\label{heff}
\eeq 
Although $H \simeq 0.9G$ alone is not enough to produce the diquark BEC (see Fig.~\ref{fig:njl0}),
the chiral-diquark coupling $K'$  increases $H'$ sufficiently for a BEC  to develop
 (see Fig.~\ref{fig:njl1}).  

    Analytically, the distinction between the BEC and BCS regimes lies 
in the nature of the quasiparticle dispersion relations, Eqs.~(\ref{eq:dispersion1}) 
and (\ref{eq:dispersion2}).  For $\mu>M$, the minima of the dispersion relations 
are at nonzero momentum $p=\sqrt{\mu^2 - M^2}$, with excitation gaps $\Delta_1$ 
and  $\Delta_8$, a structure characteristic of the BCS regime.  
On the other hand, for $\mu<M$, the minima of the dispersion curves are at $p=0$, 
a structure characteristic of the BEC regime \cite{Leggett1980}.
Figures~\ref{fig:njl1} (a) and (b) show the curve $\mu=M(\mu,T)$ as the dotted line 
in the COE region. A BEC of bound diquarks exists  between the solid and 
dotted lines.  (Note that at $T=0$ the dotted line ends at
$\mu$ = 286.6 MeV in case I
 and at $\mu$= 297.8 MeV in case II, reflecting
 the decrease of $M(\mu,T=0)$  from its vacuum value, Table~\ref{tab}.)

The structure of the crossover from BEC to BCS, at the NG-BEC and NOR-COE boundaries
 in Figure~\ref{fig:njl1},
is most clearly defined in terms of the diquark  correlation function
 \beq
G_D(\tau,{\bf x}) \equiv -4H^2 \langle T_\tau 
[s_{A}(\tau,{\bf x})s^\dagger_{A}(0,{\bf 0})]
\rangle,
\eeq
with $s_{A}(\tau,{\bf x})=q^T(\tau,{\bf x})C\gamma_5\tau_A\lambda_{A}q(\tau,{\bf x})$ 
(no summation over $A$).
In the random phase approximation (RPA), this correlation function in the complex frequency 
($z$) plane at temperatures above the  diquark condensation temperature $T_c$ is given by
\begin{equation}
  G_D^{-1}(z,{\bf q=0})=\frac{1}{4H'}
 - 4\sum_{\mp}\int_{p \le \Lambda} \frac{d^3p}{(2\pi)^3}\frac{1-2f(E_p \mp \mu)}
 {2 (E_p \mp \mu) \mp z},
\label{gdrpa}
\end{equation}
where $f(\epsilon)=1/(e^{\epsilon/T}+1)$ is the Fermi distribution
function and $\Lambda$ is the ultraviolet cutoff.  As we see from
Eq.~(\ref{gdrpa}),  $G_D$ has a branch cut on the real axis for $z \ge
2(M - \mu)$ (as well as a branch cut for  $z \le -2(M+\mu)$ from the
antiparticle contribution).   

   In the regime $\mu < M(\mu,T)$, for sufficiently large $H'$, $G_D(z,{\bf 0})$ has a pole 
on the real frequency axis for $0 \le z = M_D(\mu,T) - 2\mu \le 2(M-\mu)$, 
with $M_D(\mu,T)$ the mass of a bound diquark.  The system undergoes a BEC condensation when, 
at a given temperature, this pole first reaches zero frequency, $G_D^{-1}(0,{\bf 0})=0$ 
\cite{Nozieres1985,bbhlv,Nishida2005}; thus the condition
\begin{eqnarray}
2 \mu = M_D(\mu,T),
\end{eqnarray}  
determines the NG-BEC boundary (Fig.~\ref{fig:njl1}). 
With increasing $\mu$ the branch cut starting at $2(M-\mu)$ 
eventually reaches down to $z=0$, at which point the pole at the origin begins to 
move to complex values in the second Riemann sheet. For $\mu > M(\mu,T)$,  
the condition $G_D^{-1}(0,{\bf 0}) = 0$ defines
the onset of BCS pairing, studied in detail in Ref.~\cite{Kitazawa:2001ft}, 
and determines the NOR-COE boundary (Fig.~\ref{fig:njl1}). 
   
\begin{figure}[t]
\begin{center}
\includegraphics[width=8cm]{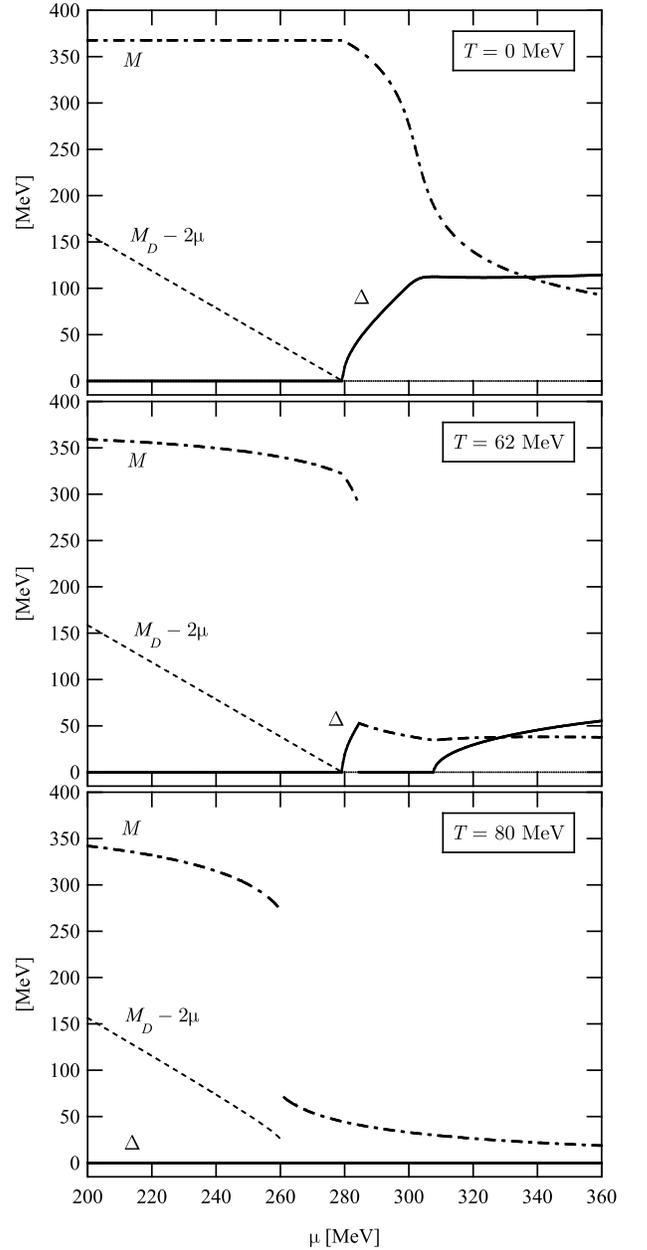}  
\caption{The Dirac mass ($M$) and
the Majorana mass ($\Delta$)  as functions of $\mu$ for $T=0$ MeV
 (top), $62$ MeV (middle), and $80$ MeV (bottom).
The excitation gap of the bound diquark in the medium, $M_D-2\mu$,
 extracted from the isolated zero of Eq.~(\ref{gdrpa}), is also shown.
The parameters are the same as in Fig.~\ref{fig:njl1}(b). }
\label{fig:Massesvsmu}
\end{center}
\end{figure}

 \begin{figure}[t]
\begin{center}
\includegraphics[width=7.5cm]{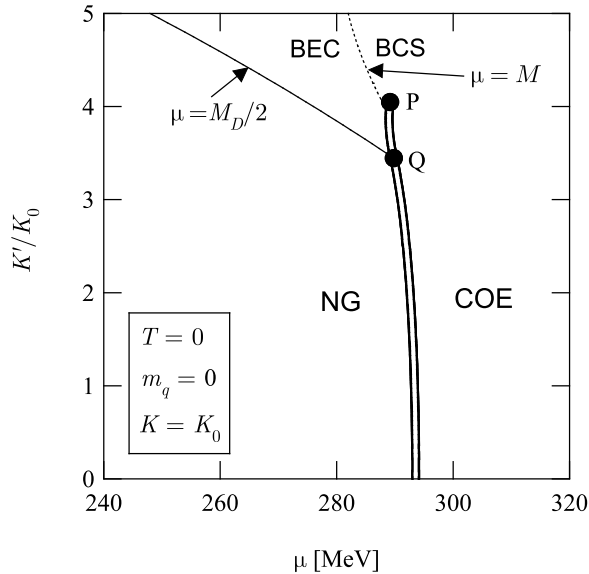} 
\caption{The phase diagram in the $(\mu,K')$-plane at $T=0$ for massless quarks, with 
the NG and COE phases. The BEC-BCS crossover in the COE phase for large $K'$ is shown 
as a dotted line. The critical point and the critical end point are denoted 
by P and Q, respectively.}
\label{fig:muKprime}
\end{center}
\end{figure}

Due to the abrupt change of $\chi$ across
the  first-order transition line,
 the NG-BEC  and NOR-COE  boundaries are not 
 smoothly connected; the former 
 touches the first-order line at higher
 temperature than the latter (see Fig.~\ref{fig:njl1}).
  As discussed more fully in \cite{Yamamoto2007a}, 
  this difference can be understood by noting that a larger chiral condensate $\chi$
  reduces the density of states at the Fermi surface and simultaneously
  increases the effective $qq$ coupling $H'$  [see Eq.~(\ref{heff})].
  Since the latter effect dominates in the present parameter set,
    the critical temperature for  
   diquark pairing is larger on the left side of the double line.   
   
The behavior of the Dirac mass $M$, the Majorana mass or gap $\Delta$, and the bound 
diquark mass $M_D$ associated with the phase diagram in Fig.~\ref{fig:njl1}(b),
are plotted  in Fig.~\ref{fig:Massesvsmu}  as functions of  $\mu$ for three temperatures, 
$T$=0, 62, and 80 MeV. At zero temperature, we see the successive transitions from
the NG phase, the BEC regime in the COE phase, to the BCS regime in the COE phase, 
with increasing $\mu$. 
The onset of BEC at $\mu=279.2$ MeV is determined by the condition $M_D-2\mu=0$.
For $T=62$ MeV, we see rather the successive transitions from the NG phase, 
the BEC regime in the COE phase, the NOR phase, to the BCS regime in the COE phase, 
with increasing $\mu$.  
The transition from the BEC regime in the COE phase to the NOR phase is first-order 
at $\mu=284.5$ MeV, with both $M$ and $\Delta$ jumping discontinuously.
The two phase transitions, from the NG phase to the BEC regime in the COE phase, 
and from the NOR phase to the BCS regime in the COE phase, signal the onset of 
a non-zero Majorana mass.  
Both transitions are correctly described in terms of the diquark correlation function 
[see Eq.~(\ref{gdrpa})]. For $T=80$ MeV, the system undergoes a first-order transition, 
from the NG phase to the NOR phase at $\mu=260.7$ MeV, 
which takes place before bound diquarks start to condense. 

Finally we consider the effect of the chiral-diquark coupling $K'$ 
 on the phases at $T=0$. 
Figure~\ref{fig:muKprime} shows the phase diagram in the
 ($\mu,K'$)-plane for massless quarks. (A similar structure holds for
 finite mass quarks.)  For small $K'$, the system has an NG phase and a
 COE phase separated by a first-order line indicated by the double line
 which eventually terminates for large $K'$ at the critical point P. On
 the other hand, for $K'$ sufficiently large compared with the cubic
 coupling, $K$, of the chiral field, a BEC regime of bound diquarks
 appears across a second-order phase transition (solid line) from the NG
 phase at a critical chemical potential $\mu=M_D/2$; the phase boundary
 joins the first-order line at the critical end point Q.  The dotted
 line, $\mu = M(\mu,T)$, shows the BEC-BCS crossover; for somewhat
 smaller $K'$, a novel first-order transition from the BEC to BCS regimes
 appears between P and Q, with discontinuous changes of both the chiral
 and diquark condensates.

\section{Discussion}
\label{sec:summary}
We have explored here the phase structure of dense three-flavor
matter using the Nambu--Jona-Lasinio model incorporating the attraction 
between the chiral and diquark condensates induced by the axial anomaly.
We demonstrated that the low temperature critical point between the hadronic phase and 
the color superconducting phase predicted by the previous 
Ginzburg-Landau analysis \cite{Hatsuda:2006ps, Yamamoto2007a} 
indeed appears in the phase diagram for sufficiently large chiral-diquark coupling.
We have also shown in Eq.~(\ref{heff}) that the axial anomaly enhances the attractive 
interaction between quarks, leading to the emergence of a Bose-Einstein
condensate (BEC) of diquark molecules. As a result, a BEC-BCS crossover 
in the diquark pairing appears in the coexistence phase, which has both 
non-zero chiral and diquark condensates.

In the phase diagram of the NJL model derived here, the BEC regime is realized 
adjacent to the lower density Nambu-Goldstone phase of massive quarks.
In QCD, however, the low density phase is in reality nuclear matter;
There remains the important problem of learning how the gas of bound diquarks 
and unpaired quarks undergoes a transition to a gas of three-quark bound states, 
or nucleons, at low density.  
Describing this transition will require going beyond the mean-field
approximation, and RPA, to take into account residual interactions
between the diquarks and unpaired quarks.   
Recent work \cite{Maeda2009,Hatsuda2009} on mixtures of bosonic and fermionic atoms 
indicates a phase diagram very reminiscent of this scenario in QCD.

It is also important to make our phase diagram more realistic by including effects 
such as the Fermi momentum mismatch induced by a strange quark mass, 
charge neutrality and $\beta$-equilibrium.
Open questions include whether the low temperature  critical point can survive 
in an inhomogeneous chiral crystalline phase \cite{Nakano2005, Nickel2009} or 
the Fulde-Ferrell-Larkin-Ovchinnikov phase
\cite{Alford2001, Bowers2002, Casalbuoni2005, Mannarelli2006, Rajagopal2006, Nickel2009a},
and how the phase structure obtained here is affected by the confinement-deconfinement 
phase transition characterized by the Polyakov loop 
\cite{Fukushima2004a,*Fukushima2008b, Ratti2006,Roessner2007,Hell2009}.
We defer these problems to future publications.

\

\

\begin{acknowledgments}
We would like to thank S.~R\"{o}ssner and W.~Weise for useful discussions.
A part of numerical calculations was carried out on Altix3700 BX2 at 
YITP in Kyoto University.
The work of GB  was
supported in part by NSF Grant No.~PHY07-01611; he also
 thanks the G-COE program of the Physics Department of the 
University of Tokyo for hospitality and support during the completion of 
this work. TH was supported in part by the Grant-in-Aid for Scientific 
Research on Innovative Areas (No. 2004: 20105003).
NY is supported by the Japan Society for the Promotion of Science for
Young Scientists. 
\end{acknowledgments}

\end{document}